\documentstyle[preprint,revtex]{aps}
\tightenlines
\begin{document}
\draft

\begin{title}
Electrostatic screening in fullerene molecules.
\end{title}

\author{J. Gonz\'alez \\}
\begin{instit}
Instituto de Estructura de la Materia. \\
Consejo Superior de Investigaciones Cient{\'\i}ficas. \\
Serrano 123, 28006-Madrid. Spain.
\end{instit}

\author{F. Guinea \\}
\begin{instit}
Instituto de Ciencia de Materiales. \\
Consejo Superior de Investigaciones Cient{\'\i}ficas. \\
Cantoblanco. 28049-Madrid. Spain.
\end{instit}

\author{M. A. H. Vozmediano \\}
\begin{instit}
Departamento de F{\'\i}sica Te\'orica. \\
Facultad de Ciencias F{\'\i}sicas.\\
Universidad Complutense.\\
Avda. Complutense s/n 28040 Madrid. Spain.
\end{instit}

\receipt{}

\begin{abstract}

The screening properties of fullerene molecules are described
by means of a continuum model which uses the electronic wavefunctions
of planar graphite as a starting point.
The long distance behavior of the system
gives rise to a renormalizable theory, which flows towards
a non trivial fixed point. Its existence implies an anomalous
dielectric constant. The screening properties
are neither metallic nor insulating.
Alternatively, the intramolecular screening
is obtained from a simple approximation to the electronic
wavefunctions. Intermolecular effects are also calculated,
As a consistency check, it is shown that the observed
polarizability of C$_{60}$ is well eproduced.

\end{abstract}

\pacs{75.10.Jm, 75.10.Lp, 75.30.Ds.}

\narrowtext

The localization of the electrons within the C$_{60}$ spheres
implies that electrostatic interactions cannot be neglected
in fullerene crystals. The simplest estimate for the charging
energy of a C$_{60}$ molecule, $e^2 / R$, where $R$ is the radius,
gives a rather large value, $\approx 4$eV. The intermediate size
of the molecule puts it in a regime different from
an isolated atom or a bulk system. Different approaches have
been used in the study of electrostatic effects
\cite{Ritchie,Auer,Dressel,Gunnarson}.
It has also been proposed that they can be the
origin of superconductivity in doped systems\cite{Kivelson,TDLee}.

In the present work, we use a simple model for the
long wavelength electronic properties of fullerene
molecules\cite{PRL,NuclP}, to analyze the screening properties.
It has been shown that it gives a reasonable
approximation to the electronic levels of C$_{60}$. Furthermore,
the scheme is sufficiently general to describe
other systems with similar topology, like nanotubes and
fullerene 'onions'. As shown in a previous work\cite{PRB},
the model can be used to study other features, like the
electron-phonon interaction' with reasonable
accuracy. Calculations based on this model
\cite{PRB} for the electrostatic interactions between the
highest occupied orbitals (in doped systems) are in good
agreement with more detailed calculations\cite{TDLee}.

The scheme is based on the fact that the low lying
electronic states of an isolated graphitic sheet are
well approximated by an effective Dirac equation in
(2+1) dimensions, using H\"uckel's theory of conjugated
carbon compounds. This description plays the same role as
the effective mass theories used in the long
wavelength regime of other materials. The existence
of rings with an odd number of atoms in C$_{60}$ leads
to additional complications, which require the
introduction of a fictitious gauge field, to account for
the frustration that these rings induce in the electronic
states.

We use the preceding scheme to analyze first the
screening properties of an isolated graphite plane.
The fact that the dielectric constant is anomalous can be
inferred from the semimetallic nature of the system. The
density of states at the Fermi level is zero, and there
is no Thomas Fermi screening at long distances. On the
other hand, there is no energy gap, so that the system
does not behave like an insulator. We can write
an effective action for the system, coupling the
Dirac electrons to the electromagnetic field in
the standard way:

\begin{equation}
{\cal S} = \int d^{2}r dt \; \overline{\Psi }(\gamma_{0}
\partial_{0} + v_{F} \mbox{\boldmath $\gamma \cdot \partial$ } )
\Psi  - i \: \frac{e}{c} \int d^{2}r dt \; \overline{\Psi }(\gamma_{0}
A_{0} + v_{F} \mbox{\boldmath $\gamma \cdot A$ } ) \Psi
\end{equation}
The action is {\em not} Lorentz invariant, as the velocity of
propagation of the fermions, $v_{F}$, is different (and much
smaller) from that of the photons, $c$. Moreover, note that the
Fermi fields are defined in  a two  dimensional plane, while the
vector potential exists throughout the entire three dimensional
space. In this way we ensure the  correct interaction at long
distances  between charges  in the plane,
$e^2  / | {\bf r} - {\bf r'} |$ .
A  consistent coupling of the electromagnetic field
is attained, in the  quantum theory, by performing the gauge
fixing so that the spatial component transverse to the plane
decouples from the  electronic system.  This is  achieved
precisely by the Feynman  gauge, in which $A_{\mu}$ has the same
direction as the current and each
component does not mix with the rest
\begin{equation}
\langle T A_{\mu}(t, {\bf r}) A_{\nu}(t', {\bf r}') \rangle =
- i \: \delta_{\mu \nu} \int \frac{d^{3}k}{(2 \pi)^{3}} \frac{d
\omega}{2 \pi}  \; \frac{\mbox{\Large $e^{i {\bf k} \cdot ({\bf r} -
{\bf r}') - i \omega (t - t') }$ } }{k^{2} - \omega^{2} - i \epsilon}
\end{equation}

{}From dimensional counting, it is easy to conclude that the
coupling constant $e$ is dimensionless (in units in which
$\hbar = c = 1$), so that perturbation
theory should give rise to logarithmic corrections, which can
be included in renormalized parameters in the standard way.
We are interested in the low energy properties, which are given
by the infrared fixed point of the theory. Our initial conditions
are fixed by the known bare parameters, $v_F$ and $e$ at the
upper cutoff where our continuum model is valid. This cutoff,
in energies, is determined by the bandwidth of the $\pi$
electron band that we are considering, which is proportional,
within H\"uckel theory, to the interatomic hopping matrix element
of the hamiltonian, $t \approx 2.2$eV\cite{Huckel}. In space,
the relevant cutoff is the interatomic distance. $a = 1.4$\AA.
In terms of these parameters, $v_F = 3 t a / (2 \hbar)$.

Regularizing the theory by analytic continuation to dimension $d
= 3 - \varepsilon$\cite{ramond}, the divergent contributions to the quantum
effective  action are given, to lowest order in  $e^{2}/\hbar
c$, by the diagrams in Figure 1. It can be shown that the  two
dimensional matter fields do not renormalize the scale of the
electromagnetic  field. The self-energy contribution from Figure
1a gives  rise to the renormalization of $v_{F}$, $v_{F} = Z_{v_F}
(v_{F})_{R}$, as well as to the wavefunction renormalization,
$\Psi = Z_{\Psi} \Psi_{R}$. In terms of hypergeometic functions,
we obtain:

\begin{eqnarray}
Z_{v_F} & = & 1 - \frac{1}{4 \pi} \frac{e^{2}}{\hbar c}
 \left\{ \pi \frac{c}{v_{F}} F \left(\frac{1}{2}, \frac{3}{2}; \frac{1}{2};
 \frac{v_{F}^{2}}{c^{2}} \right) - 2 \pi \frac{v_{F}}{c} \left( 1 - 2
\frac{v_{F}^{2}}{c^{2}} \right)
   F \left(\frac{3}{2}, \frac{3}{2}; \frac{3}{2};
 \frac{v_{F}^{2}}{c^{2}} \right)   \right.  \nonumber   \\
      &   &    \left. + 4 \left( 1 - 2
\frac{v_{F}^{2}}{c^{2}} \right) F \left(1, 1; \frac{1}{2};
 \frac{v_{F}^{2}}{c^{2}} \right) -
4 F \left(1, 2; \frac{3}{2};
 \frac{v_{F}^{2}}{c^{2}} \right)  \right\} \frac{1}{\varepsilon}
  +  O   \left(\left(\frac{e^{2}}{\hbar c}\right)^{2}
                                    \right)
                                           \nonumber     \\
Z_{\Psi} & = & 1 + \frac{1}{2 \pi} \frac{e^{2}}{\hbar c}
\left( 1 - 2 \frac{v_{F}^{2}}{c^{2}}  \right)
 \left\{ 2 F \left(1, 1; \frac{1}{2};
 \frac{v_{F}^{2}}{c^{2}} \right) - \right. \nonumber \\
& & \left. \pi \frac{v_{F}}{c}
 F \left(\frac{3}{2}, \frac{3}{2}; \frac{3}{2};
 \frac{v_{F}^{2}}{c^{2}} \right) \right\}
\frac{1}{\varepsilon}
  +   O   \left(\left(\frac{e^{2}}{\hbar c}\right)^{2}
                                    \right)
\end{eqnarray}
The correction to the vertex from Figure 1b gives the
renormalization of the electric charge, $e = Z_{e} e_{R}$, and
provides at the same time an alternative way of computing the
renormalization of $v_{F}$. The value of $Z_{v_F}$ from the vertex
correction coincides precisely with the above expression from
the self-energy diagram, ensuring the renormalizability of
the model. The second remarkable point is that, to this
perturbative order, $Z_{e}$ turns out to be equal to $1$. This
means  that both pieces of the action (1) are affected by
the same global renormalizatiton and that, therefore, the gauge
invariance of the theory is maintained at the quantum level.

In the relevant physical limit, $v_F / c \rightarrow 0$,
the structure of (3) is greatly simplified, and we obtain:

\begin{eqnarray}
Z_{v_F} & = & 1 - \frac{1}{4} \frac{e^{2}}{\hbar v_{F}}
  \frac{1}{\varepsilon} + O
\left(\left(\frac{e^{2}}{\hbar v_{F}}\right)^{2}
                                    \right)    \nonumber \\
Z_{\Psi} & = & 1 + O
  \left(\left(\frac{e^{2}}{\hbar v_{F}}\right)^{2}
                                    \right)
\end{eqnarray}

Note that the expansion, which originally was on the
fine structure constant, $e^2 / ( \hbar c )$ is changed
into an expansion on $e^2 / ( \hbar v_F ) \propto
e^2 / ( t a )$. This was to be expected on physical grounds, as
in the limit when the propagation of the electromagnetic field is nearly
instantaneous, this is the only measure of the
relative strength of the electrostatic and kinetic
energies.
This parameter is greater
than unity, although the convergence of the series is improved
by the presence of dimensionless constants in the
denominators of all terms.
The fact that the electric charge is not renormalized
is true to all orders, in the limit $v_F / c \rightarrow 0$.
The only coupling which remains is longitudinal, and the
matrix elements for the relevant transitions tend to zero
for small momentum transfers.

Applying standard renormalization group methods\cite{ramond}, we obtain the
variation of the renormalized coupling $ta/e^{2}$ as  a function
of the energy scale $\mu$

\begin{equation}
\mu \frac{d}{d\mu} \left( \frac{ta}{e^{2}} \right) =
    - \frac{1}{4} +
      O \left( \frac{e^{2}}{ta}  \right)
\end{equation}

The computation of higher perturbative orders simply amounts to
adding higher powers of $e^{2}/(ta)$ to the right hand side of
the equation, and does not alter the qualitative behaviour of
the renormalized coupling. From (5) it is clear, in fact, that
perturbation theory becomes more accurate as the spatial scale
increases in the infrared regime.

The scaling of $v_F$ changes the density of states
al low energies, which behaves as:

\begin{equation}
N ( \epsilon ) \sim | \epsilon |^{1 + e^2 / ( 4 t a ) +
O ( e^4 / ( t a )^2 )}
\end{equation}

The scaling outlined above is limited, in a fullerene
sphere, by its radius, which acts as an infrared cutoff.
In a doped compound, another low energy scale arises:
the Fermi energy associated with the extra electrons,
or a length $\propto k_F^{-1}$. In the cases of interest,
the filling factor is small, 3/60, so that the relevant
cutoff remains the radius of the molecule ($k_F^{-1}
\sim 20 a$). Note also that, in doped systems,
the existence of a finite density of states at the Fermi
level allows us to define a Fermi-Thomas screening
length ( $k_{FT}^2 \sim e^2 k_F / ( t a R )$ ). Again,
for the relevant values of $k_F$, this length is greater
than the radius. Hence, metallic screening plays no role
on the intramolecular screening in doped C$_{60}$.

The absence of renormalization of $e^2$,
mentioned earlier, implies that the charging energies
scale as $e^2 / R$, although with a proportionality term
reduced by the finite corrections.

We can also use our continuum aprroximation to the
electronic structure of the fullerenes to study their
electrostatic properties in a different way to the one
discussed before. In principle, the knowledge of
all the (discrete) electronic levels, allows us to
use standard second order perturbation theory, and
calculate the polarization of the inner levels as
fucntion of the extra electrons added to the molecule.
In this way we can estimate the attractive interaction
between two additional electrons, induced by the
polarization of the other levels. The level spacings
scale like $t a / R$, so that, on dimensional grounds,
this interaction behaves as $e^4 / ( t a R )$, while the
direct Coulomb repulsion is $\propto e^2 / R$.
Thus, the balance
between the two depends crucially on the dimensionless
prefactor. For the wavefunctions of the lowest
unoccupied triplet, this value is given by a
convergent sum, which involves transitions from
multiplets which are
higher and higher in energy. The convergence of this
sum has the same origin as the finiteness of the
diagrammatic corrections to $e^2$ in the previous analysis.
{}From the computation of this series, we find a value
of 0.14. Hence, the polarization of the inner levels
in C$_{60}$ reduces by a significant factor $\sim$
50 \% the bare Coulomb interactions. We can also
compute the effective interaction between two
additional electrons in the lowest triplet. This
interaction is attractive, and reduces strongly,
but does not cancel, the bare repulsion.

Finally, we can compute the interaction between
electrons located in different molecules.
For two large molecules at close distance, the
most convinient starting point is tha analysis
of two planes of graphite. The screening of
the electrostatic interactions is
described by the scaling given in eq. (5),
until a scale proportional to the interplane distance is
reached. Then, additional screening is induced
by the mutual polarization effects of the two
planes. This effects leads to an interaction, which,
to lowest order, goes like $e^4 \times  A / ( t a D^3 )$,
where $A$ is the area of the planes, and $D$ the interplane
distance.

At sufficiently large distances, each molecule is
described
by its polarizabilty in a standard way. It can be computed
from the knowledge of the electronic transition energies,
and the corresponding matrix elements. For a
neutral, spherical fullerene, we find:

\begin{equation}
P = {{e^2 R^3}\over{ t a}} \left[ {1\over{3}} +
\sum_{l = 2}^{\infty} {{8 l ( l + 2 )}\over {9 ( l + 1 )
\sqrt{2 ( l^2 + 2 l - 1 )}}} \left(
1 - \sqrt{{{(l - 1 ) ( l + 1 )}\over{( l + 1 ) ( l + 2)}}}\right)^2
\right]
\end{equation}

Inserting the appropiate parameters for C$_{60}$, we obtain
$P = 90$\AA, which is in very good agreement with the
experimental value, 80 \AA \cite{Ritchie,pol}, as well
as with more sophisticated theoretical calculations
\cite{Gunnarson,Shuai,Quong}.

{}From the scaling of the density of states (6), we can also infer the
role of intermolecular tunneling. The anomalous terms
in the exponent reduce the effective tunneling at low energies,
in a similar way to the case of a 1D Luttinger
liquid\cite{Wen,Schultz,GZ}. If the correction to the
non interacting exponent is greater than one, it can be
shown that intermolecular tunneling becomes
an irrelevant perturbation at low energies. In the present situation,
that is not tha case. However, the surviving effective
tunneling should be greatly reduced with respect to its
bare value \cite{GZ}:

\begin{equation}
t'_{ren} \sim t'_{bare} \left( {{t'_{bare}}\over{t}}
\right)^{{e^2}\over{4 t a}}
\end{equation}

In summary, we have studied the screening properties
of fullerene molecules. The perfect semimetallic
nature of isolated graphite planes
leads to unusual features, like the absence
of metallic screening, although the system does not
behave like an insulator.  There is a significant reduction of
the intramolecular charging energies, and an associated
reduction in the effective intermolecular hopping.
We have also shown how to compute the most relevant intermolecular
electrostatic effects.

{}From a fundamental point of view, we have found that the
low energy properties of the system give rise to a
non trivial, renormalizable, field theory,
which looks like a simplified version of QED. The anomalous
dimensions acquired by the propagators resemble, in many ways,
the one dimensional Luttinger liquids.

\figure{ Low order diagrams computed in the text. a) Self energy
correction. b) Vertex correction. The convention used for
the momenta and energies is that of the limit $v_F / c \rightarrow 0$.}

\end{document}